\documentclass[prl,twocolumn,reprint]{revtex4-1}

\usepackage[english]{babel}
\usepackage{graphicx}
\usepackage{epstopdf}

\begin{document}

\title{Using technical noise to increase the signal-to-noise ratio, via imaginary weak values. }

\author{Y. Kedem}
\affiliation{ Raymond and Beverly Sackler School of Physics and Astronomy\\
 Tel-Aviv University, Tel-Aviv 69978, Israel}

\begin{abstract}
The advantages of weak measurements, and especially measurements of imaginary weak values, for precision enhancement, are discussed. A situation is considered in which the initial state of the measurement device varies randomly on each run, and is shown to be in fact beneficial when imaginary weak values are used. The result is supported by numerical calculation and also provides an explanation for the reduction of technical noise in some recent experimental results. A connection to quantum metrology formalism is made.
\end{abstract}

\maketitle

In 1988 Aharonov, Albert and Vaidman (AAV) \cite{AAV88} discovered that the measured value of an observable can be 100 times bigger than its biggest eigenvalue, provided the measurement interaction is weak and a postselection is employed.  They showed that a system which is coupled weakly to another, pre- and postselected system, described by the
 {\em two-state vector} $\langle \Phi |~|\Psi\rangle $, via an observable $C$, is effectively coupled to the weak value of the observable  \cite{AV90}
\begin{equation}\label{wv}
C_w \equiv { \langle{\Phi} \vert C \vert\Psi\rangle \over
\langle{\Phi}\vert{\Psi}\rangle } .
\end{equation}
The replacement of the interaction operator with its weak value, which is a  complex number \cite{Joz}, is known as the AAV effect and the procedure in which the weak value is measured is referred to as a weak measurement. The promise that this phenomenon holds for improving precision measurements had recently started to materialize in observation of the spin hall effect of light  \cite{HK} and ultra sensitive measurement of beam deflection \cite{How}. Other areas where the use of weak measurements was investigated include measuring small longitudinal phase shifts  \cite{Brunner,phase}, charge sensing  \cite{charge}, frequency measurements \cite{freq}, and Kerr nonlinearities \cite{stein}. 

The general use of quantum effects for precision enhancements, known as quantum metrology \cite{metro}, is showing significant results \cite{inter} and lately, much attention is drawn to practical issues such as the effects of an environment \cite{env, env2},  noise \cite{esch},  and technical limitations \cite{real}. According to \cite{HK,How} the use of imaginary weak values in the measurement process, allows a reduction in technical noise. In this letter we will analyze the process of weak measurement as a method for precision measurements. Furthermore, we will present a concrete model for technical noise affecting the preparation of the measurement device (meter), and show that in the presence of such a noise the precision is enhanced. 

We start with an overview of known results regarding the precision achievable by weak measurements. Consider a physical interaction:
\begin{equation}\label{h}
   H = g(t) PC ,
\end{equation}
where $C$ is an observable on a system, $P$ is an operator on a meter and $g(t)$ is a coupling function satisfying $ \int g(t) dt =k$. Our concern is estimating the size of $k$, or in some cases simply observing the interaction. A straight forward approach is to put the system in an eigenstate of $C$ having some eigenvalue $c$, and the meter in a Gaussian state:
\begin{equation}\label{Pgauss}
 \Psi_{M} (Q)  =(\Delta ^2 \pi )^{-1/4}  e^{ -{Q ^2 \over 2 \Delta ^2}},
\end{equation}
where $Q$ is a variable conjugate to $P$, and $\Delta$ is its quantum uncertainty. An estimate of $k$  can be obtained from the shift in $Q$ due to the interaction, $\langle Q \rangle = kc$ , and its precision is determined by the standard deviation ${1 \over \sqrt{2}} \Delta.$ In the case $kc  \ll \Delta$, little information is acquired from a single measurement, but by repeating the procedure $N$ times and averaging the results, the precision is enhanced. Strictly speaking, the amount of information gathered, regarding $k$, is measured by the Fisher information \cite{esch}, but for our purposes we can use the more intuitive concept of {\em signal to noise ratio} ($S/N$) \cite{snr}, which in this case is
\begin{equation}\label{snr}
S/N = \sqrt{N }{kc   \over \Delta }.
\end{equation}

Since our interest is in the regime where  $kc  \ll \Delta$, which is a condition for the AAV effect \cite{wu}, we will, for now, assume that the AAV effect occurs and later examine its validity in more detail. Thus, we will consider the system to be initially in a state $|\Psi\rangle$ and take into account the meter results only when the system was found in a state $|\Phi\rangle$, after the interaction, which implies a replacement $C \rightarrow C_w $ in (\ref{h}) \cite{KV}. The shift in $Q$ is given by $\langle Q \rangle_{\Phi}  = k ReC_w$ \cite{AAV88}, and
\begin{equation}\label{snr2}
S/N = \sqrt{N_{\Phi} }{k {\text Re}C_w \over \Delta },
\end{equation}
 where $N_{\Phi} \sim N \left| \langle \Phi |\Psi\rangle \right|^2$ is the number of times the system was found in a state  $|\Phi\rangle$. In order for  $C_w$ to be larger than any eigenvalue of $C$, the scalar product, $\langle \Phi |\Psi\rangle,$ has to be small, so we can see that we cannot improve (\ref{snr2}) significantly, relative to  (\ref{snr}). It is, however, a remarkable fact that by using only a small portion of our potential data, we get the same quality of information. In practice, there are many set ups where a rare postselection is beneficial, especially when there is a detection constraint, such as saturation limits or dead time.

Another option is to measure the meter in the $P$ basis. Assuming the meter initial state is (\ref{Pgauss}), which we can write in the $P$ basis as: $ \Psi_{M} (P)  = (\Delta ^{-2} \pi )^{-1/4}  e^{ -{\Delta ^2 P ^2 \over 2 }}$,
the final shift in $P$ is given by $\langle P \rangle_{\Phi}  =  k \Delta^{-2}{\text Im} C_w$ \cite{Joz} and the standard deviation is ${1 \over \sqrt{2}} \Delta^{-1}$, giving us
\begin{equation}\label{snr3}
S/N = \sqrt{N_{\Phi} }{k {\text Im} C_w \over \Delta }.
\end{equation}
Surprisingly, for ${\text Im} C_w = {\text Re} C_w$, the S/N for this case is the same as (\ref{snr2}) and it seems measuring an imaginary weak value is ineffective. However, as we will now show, this is not the case.

In calculating the $S/N$ (\ref{snr}),  (\ref{snr2}) and  (\ref{snr3}) we considered only the quantum uncertainty, sometimes called shot noise, and not any technical issues. Since the set ups used in advance experiments are highly intricate, there is an enormous range of possible technical issues and conceiving a general model for their effect is beyond the scope of this letter. Instead, we will restrict our discussion to faults in the preparation of the meter, causing its initial state to be shifted with respect to  (\ref{Pgauss}). 

Let us start by considering a shift, $Q_0$, in the $Q$  basis only, making the initial state of the meter  
\begin{equation}\label{gaussq}
 \Psi_{M} (Q)  =(\Delta ^2 \pi )^{-1/4}  e^{ -{(Q-Q_0) ^2 \over 2 \Delta ^2}}.
\end{equation}
A measurement of $Q$, after an interaction (\ref{h}) with a pre and postselected system $ \langle \Phi |~~|\Psi\rangle $, will yield
\begin{eqnarray}\label{Q1}
  \langle Q \rangle_{\Phi} &=& Q_0 + k {\text Re}C_w,  \nonumber \\
 \langle Q^2 \rangle_{\Phi} &=& {\Delta ^2 \over 2}+ (Q_0 + k{\text Re}C_w)^2. 
\end{eqnarray}
Since the shift  $Q_0$ can be different for every run, some distribution should be used when averaging over the results. We assume an uncorrelated distribution with vanishing average  $ \overline{Q_0}=0$, which can be seen as white noise. A finite average would describe a systematic error while correlations can appear, for example, if $Q_0$ has some time dependency which is relevant to the frequency in which the runs occur or to their total time. In order to treat such disturbances, an analysis using Allan variance \cite{allan} is needed which we will not discuss here. In \cite{stein}, weak measurements was shown to be beneficial for noise with long correlation time, however, their results about its ineffectiveness for white noise was based on the measurements of real weak values. 

We consider the probability of a shift $Q_0$ to be
\begin{equation}\label{prob}
\Pr(Q_0)= (\Delta_Q \sqrt {\pi} )^{-1} e^{ -{Q_0 ^2 \over \Delta_Q ^2}},
\end{equation}
where $ \Delta_Q$ is the width of the distribution of the shift. The only essential characteristics of the distribution, to our results, are $ \overline{ Q_0 } = 0$ and $ \overline{ Q_0 ^2 } =  { \Delta_Q ^2 \over 2} $, so taking it to be a Gaussian is for strictly for the simplicity of presentation. 
 An average over $Q_0$ will result in
\begin{eqnarray}\label{Q2}
 \overline{\langle Q \rangle_{\Phi} } &=& k{\text Re}C_w,  \nonumber \\
 \overline{\langle Q^2 \rangle_{\Phi} } &=& {\Delta ^2 \over 2}+ {\Delta_Q ^2 \over 2}  + (k{\text Re}C_w)^2,
\end{eqnarray}
 meaning the same shift as it was for (\ref{Pgauss}) but a larger standard deviation, making the $S/N$ smaller than  (\ref{snr2}). Similarly we can get $S/N = \sqrt{N}  k c / \sqrt{ \Delta ^2+ \Delta_Q ^2}$ if the system is in an eigenstate of $C$ with eigenvalue $c$.

By writing the meter state, in the $P$ basis, after the interaction and postselection:
\begin{equation}\label{gaussfp}
 \Psi_{M} (P)  = \mathcal{N} (\Delta ^{-2} \pi )^{-1/4}  e^{ -{\Delta ^2 P ^2 \over 2 } + i (Q_0 - k C_w) P},
\end{equation}
where $ \mathcal{N} = Exp \left[ - k^2 \Delta ^{-2} ( {\text Im} C_w) ^2 / 2  \right]$ is the re-normalization factor due to the postselection, one can see that a measurement of $P$ will yield
\begin{eqnarray}\label{P1}
  \langle P \rangle_{\Phi} &=& k \Delta ^{-2} {\text Im} C_w,  \nonumber \\
 \langle P^2 \rangle_{\Phi} &=& {\Delta ^{-2} \over 2}+ ( k  \Delta ^{-2} {\text Im} C_w )^2. 
\end{eqnarray}
 This means that the $S/N$ for this case is the same as (\ref{snr3}), the $S/N$ for the case of an ideal initial state.

This is the first result of our letter: when one has a dominant technical issue in the preparation of a variable conjugate to the interaction operator, measurements of an imaginary weak value can eliminate its effect.

Let us now consider a shift, $P_0$, in the $P$ basis, making the initial state of the meter  
\begin{equation}\label{gaussp2}
 \Psi_{M} (P)  =(\Delta ^{-2} \pi )^{-1/4}  e^{ -{ \Delta ^2 (P-P_0) ^2 \over 2}},
\end{equation}
with probability
\begin{equation}\label{prob1}
\Pr(P_0)= (\Delta_P  \sqrt {\pi} )^{-1}  e^{ -{P_0 ^2 \over  \Delta_P ^2}},
\end{equation}
where $ \Delta_P$ is the width of the distribution of the shift. After an interaction (\ref{h}) with a pre and postselected system $ \langle \Phi |~~|\Psi\rangle $ the meter is in a state 
\begin{equation}\label{gaussfp2}
 \Psi_{M} (P)  = \mathcal{N}_{P_0} (\Delta ^{-2} \pi )^{-1/4}  e^{ -{\Delta ^2 (P-P_0) ^2 \over 2 } - i k C_w P},
\end{equation}
where $\mathcal{N}_{P_0}$ is the re-normalization factor due to the postselection. A final measurement of $P$, will yield
\begin{eqnarray}\label{P2a}
  \langle P \rangle_{\Phi} &=& P_0 + k \Delta ^{-2} {\text Im} C_w,  \nonumber \\
 \langle P^2 \rangle_{\Phi} &=& {\Delta ^{-2} \over 2}+ (P_0 + k  \Delta ^{-2} {\text Im} C_w)^2. 
\end{eqnarray}
In order to calculate the average over $P_0$ we have to consider the probability of postselection
\begin{eqnarray}\label{post}
 \Pr(|\Phi\rangle \mid P_0) &=& \left|\langle\Phi\vert\Psi\rangle\right|^{2} e^{ k {\text Im} C_w \left(2 P_0 + k {\text Im} C_w \Delta^{-2}\right)}  \nonumber \\
 &=& \left|\langle\Phi\vert\Psi\rangle\right|^{2}  \mathcal{N}_{P_0}^{-2},
\end{eqnarray}
which was of no importance for a shift in $Q$, since it did not depend on $Q_0$. This means that if we prepare an ensemble of $N$ meters, with states (\ref{gaussp2}) according to the distribution  (\ref{prob1}), and then, after an interaction (\ref{h}), we postselect to $| \Phi \rangle$, the postselected ensemble of meters will have a different distribution:
\begin{eqnarray}\label{prob3}
\Pr(P_0 \mid |\Phi\rangle ) &=&  {\Pr(P_0)  \Pr(|\Phi\rangle \mid P_0) \over \Pr(|\Phi\rangle )}  \nonumber \\
 &=&   (\Delta_P  \sqrt {2 \pi} )^{-1}  e^{ -{(P_0 - k {\text Im} C_w \Delta_P ^2)^2 \over  \Delta_P ^2}}.
\end{eqnarray}
Calculating the averages using (\ref{prob3}) we get
\begin{eqnarray}\label{P2}
  \overline{ \langle P \rangle_{\Phi}} &=&  k (\Delta ^{-2}+ \Delta_P ^2)  {\text Im} C_w,  \nonumber \\
  \overline{ \langle P^2 \rangle_{\Phi}} &=&{\Delta ^{-2} \over 2}+  {\Delta_p ^2 \over 2} + \left(k (\Delta ^{-2}+ \Delta_P ^2) ImC_w \right)^2,
\end{eqnarray}
yielding:
\begin{equation}\label{snr6}
S/N = \sqrt{N_{\Phi} } k {\text Im} C_w  \sqrt{\Delta ^{-2} +  \Delta_p ^2 }.
\end{equation}
While for $ \Delta_p = 0$ this $S/N$ equals (\ref{snr3}), for $ \Delta_p > 0$  it is bigger.

This is the main result of our letter: In the regime where the AAV effect occurs, a non coherent spread in the variable appearing in the interaction improves the precision of the measurement.

Unlike $Q$, which is changed according to ${\text Re} C_w$, $P$ is a constant of motion under the Hamiltonian (\ref{h}), so the change in its distribution, can be understood via the postselection probability, $ \left| \langle{\Phi} \vert e^{-i k P C} \vert\Psi\rangle \right|^2$. This means different values of $P$ would cause different amounts of disturbance on the system and would have different probability to be found after the postselection. Expanding this probability to first order in $(k P)$: $\left| \langle{\Phi} \vert \Psi\rangle \right|^2(1+2 k {\text Im} C_w P )$, we see that, in this regime, it is indeed the imaginary part determining how the disturbance affect the probability.    

One might consider a measurement of this disturbance directly, i.e. varying $P$ and measuring the postselection probability. The binomial distribution would give an $S/N$ of $2 k {\text Im} C_w P \sqrt{N \left| \langle{\Phi} \vert\Psi\rangle \right|^2 \over (1 - \left| \langle{\Phi} \vert\Psi\rangle \right|^2) }$,
which is comparable to (\ref{snr6}) with the replacement $P \leftrightarrow \sqrt{\Delta ^{-2} + \Delta_p ^2 }$. This highlights some of the differences in the experimental challenges each method presents, with regard to the preparation and measurement of $P$. 

We turn now to examining the conditions for the AAV effect, in the context of an imperfect meter preparation. The evolution (up to normalization) is given by:
\begin{eqnarray}\label{rem}
\langle\Phi\vert e^{-i k P C}\vert\Psi\rangle &=& \langle\Phi\vert\Psi\rangle e^{-i k C_{w}P} \\
&+&\langle\Phi\vert\Psi\rangle\sum_{n=2}^{\infty}\frac{(-i k P)^{n}}{n!}[(C^{n})_{w}-(C_{w})^{n}] . \nonumber
\end{eqnarray}
The AAV effect means that the final state of the meter is determined by the first term, so we want to see when the second term is negligible. In an experiment aimed at measuring a tiny effect $k$ is extremely small, so it would be natural to look on the case where $k \rightarrow 0$ in which the condition for the AAV is trivial. 

For a more detailed condition, but one that is related to quantities which are already used, we need make some assumptions. One is that, in the sum over $n$, the first term in the sum, i.e. $n=2$, is the largest, since higher orders would be smaller. Another one is assuming $ |(C^{n})_w | < | C_w |^n $ for $n>1$, which is the case, in general, for a weak value that is larger than any eigenvalue, limiting our concern to verifying the condition: $ | k C_w | ^2 \langle P^2 \rangle \ll 1$. For the state (\ref{gaussq}), it amounts to $ |k C_w |^2 \Delta^{-2} \ll 1 $, implying that there is no dependency on the distribution of $Q_0$ and also that for a purely real (imaginary) weak value, the $S/N$ (\ref{snr2}) ( (\ref{snr3}) ) have to be small for $N_{\Phi} =1$. Thus, a small $S/N$ per measurement is a necessary condition for the AAV effect. For the state (\ref{gaussp2}), with a distribution (\ref{prob1}), we have
\begin{equation}\label{cond}
|k C_w |^2 (\Delta ^{-2} + \Delta_p ^2 ) \ll 1,
\end{equation}
implying that in order to make the S/N (\ref{snr6}) large, for any value of $ \Delta_p$, one has to perform many measurements. 

Naturally, there could be technical problems with the preparation of $|\Psi\rangle $, or the measurement of $| \Phi \rangle $, which can decrease the $S/N$. This issue is not in the scope of this paper but we can mention that since usually different experimental equipment is used for the meter and the system, for example a polarizer and a split detector, the technical problems are often not related. Furthermore, our result can assist the experimentalist in choosing what should be considered as a meter and what should be the system.

\begin{figure}
  \centering
    \includegraphics[width=0.474\textwidth]{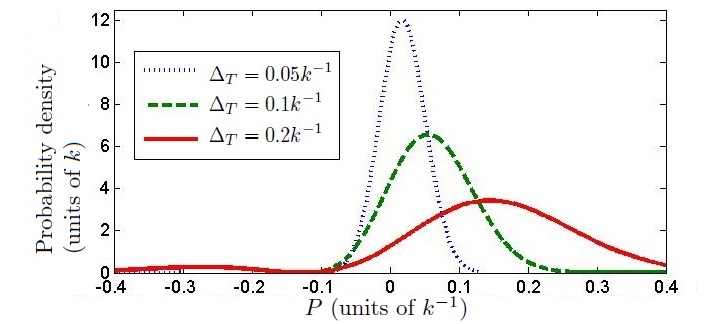}
    \includegraphics[width=0.473\textwidth]{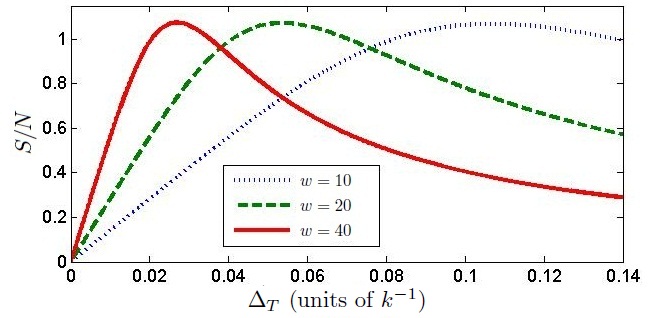}
\caption{(Color online) The expected results of a final measurement of $P$, based on equation (\ref{dist}): The probability distribution for $w = 8$ (TOP) and the $S/N$ per measurement (Bottom). For $w k \Delta_T < 1$ the form of the distribution is nearly the same and its center is shifted by $w k \Delta_T^2$, however, when $w k \Delta_T \gtrsim 1$ the form is distorted. For small $ \Delta_T$, the $S/N$ is increasing linearly, in agreement with (\ref{snr6}). The maximum is around $w k \Delta_T \sim 1$. From (\ref{cond}) it is clear that for larger values of $ \Delta_T $ the AAV effect is not valid, and thus the $S/N$ in getting smaller, as expected for standard measurements. In an experiment aimed at measuring a tiny effect, such as  \cite{HK} and \cite{How}, the interaction strength, $k$, would be very small making $ k \Delta_T \ll 1$ the relevant regime and only with the factor $ \sqrt{N_{\Phi} } \sim \sqrt{N}/w$, due to $N$ repetitions, can the $S/N$ be larger than unity. }
  \label{dist1}
\end{figure}

We support our results with a numerical calculation of a simple example, in which the pre and postselected system is a two level system (qubit) described by $(4+4w^2)^{-1/2}\left(   \left\langle \uparrow \right| +  \left\langle \downarrow \right|\right)~~\left((1+ iw) \left|\uparrow\right\rangle + (1- iw)  \left| \downarrow \right\rangle\right)$, where $ \left|\uparrow\right\rangle$ ($\left| \downarrow \right\rangle$) is an eigenstate of $C$ with eigenvalue 1 (-1).
The weak value is given by $C_w = i w$, but our calculation is not based on the AAV effect. For an initial state of the meter that is described by (\ref{gaussp2}) and (\ref{prob1}), we find that the distribution of a final measurement of $P$ is given by 
\begin{equation}\label{dist}
\rho(P)= {2e^{k^2 \Delta_T ^2 - {P^2 \over \Delta_T ^2 }} \left|\cos(k P) + w \sin(k P)\right|^2 \over \left(1 - w^2 + \left(1+w^2 \right) e^{k^2 \Delta_T ^2}\right)\sqrt{\pi} \Delta_T },
\end{equation}
where $ \Delta_T = \sqrt{\Delta ^{-2} + \Delta_p ^2}$. This distribution and the $S/N$ per measurement for it, are plotted in Fig. \ref{dist1}. 

In order to put our results in an experimental context, we analyze two experiments, \cite{HK} and \cite{How}, where weak measurements were used to detect tiny modifications in a paraxial light beam. In \cite{HK} the beam was displaced by the spin hall effect of light, creating a polarization dependent change in its transverse spatial distribution. They considered an effective Hamiltonian of the form of (\ref{h}), with $C$ being a polarization variable, $P$ the transverse momentum and $k$ was a small coefficient that needed to be estimated. Polarizers were used for the pre and postselection, making $C_w$ purely imaginary, and a position sensor was located in a distance such that the center of the spatial distribution was determined by the transverse momentum immediately after the interaction.

In \cite{How}, a Sagnac interferometer was used where the angle of one of the mirrors changed the beams direction, depending on which path it took in the interferometer. The analogue interaction of the type (\ref{h}) is having $C$ as the which-path variable, $P$ as transverse position and $k$ as the angle of the mirror times the light wave number. The interferometer was set up to make the weak value purely imaginary, and lenses were used to make the transverse position of the beam at detection proportional to the transverse position immediately after the interaction, up to a geometrical optical factor. 

Thus, even though the interactions were of different nature, both results should agree with (\ref{P2}). The manifestation of  $ \Delta ^{-2} +  \Delta_p ^2$ in an experiment would be the square of the width of the final measurement, and indeed, in both experiments the final result was proportional to this quantity. It was also mentioned in \cite{How},  \cite{HK} and  \cite{Brunner} that this method was especially beneficial for technical noise. Distinguishing between the coherent width $\Delta ^{-1}$ , and the one caused by technical issues, $ \Delta_p $, can be rather difficult, but it is unnecessary in our formalism.

Unlike the common practice in quantum metrology \cite{metro, esch}, our results do not require the meters to be entangled. The correlations created by the postselection can be viewed as classical ones and thus the precision scales as $\sqrt{N}$. Instead, the Cram\'er-Rao bound is improved simply by increasing the variance of the Hamiltonian, a task that can be done in a non-coherent way, and thus might be much simpler, experimentally, than the creation of entanglement.

Technical noise is present in any kind of experimental setup so our result can be applied to physical systems in a vast variety of fields, like solid state, optics, atomic and more.  Regarding noise as an advantage means that low-cost alternatives can be used and that elaborate noise reduction methods can be avoided. This can mean the use of white light instead of a laser  \cite{phase} or operating in room temperature and without a vacuum chamber.

We have shown that in the scenario of measurement of imaginary weak values, a shortcoming in the ability to prepare the meter in an exact known state, does not diminish the precision and the result of some flawed preparation can in fact increase the precision. This phenomenon explains some remarkable recent results where technical noise was overcome and it has the potential to improve many quantum metrology schemes in a novel way.

This work has been supported in part by  grant 
32/08 of the Binational Science Foundation, and grant
1125/10 of the Israel Science Foundationt.

\end{document}